\documentclass[conference]{IEEEtran}
\IEEEoverridecommandlockouts
\pdfoutput=1
\usepackage{amsmath,amssymb,amsfonts}
\usepackage{algorithmic}
\usepackage{graphicx}
\usepackage{textcomp}
\usepackage{xcolor}
\def\BibTeX{{\rm B\kern-.05em{\sc i\kern-.025em b}\kern-.08em
    T\kern-.1667em\lower.7ex\hbox{E`}\kern-.125emX}}

\usepackage{booktabs}					
\usepackage{multirow}                   
\usepackage[hidelinks]{hyperref}		
\usepackage{tcolorbox}					

\usepackage{amssymb}
\usepackage{color}
\usepackage{caption}                
\usepackage{soul}
\usepackage{todonotes}
\usepackage{paralist}
\usepackage[english]{babel}
\usepackage{blindtext} 
\setstcolor{red}

\usepackage{listings}

\lstdefinestyle{mystyle}{
    backgroundcolor=\color{white},   
    commentstyle=\color{orange},
    keywordstyle=\color{black}\bf,
    numberstyle=\tiny\color{magenta},
    stringstyle=\color{purple},
    identifierstyle=\color{blue},
    basicstyle=\ttfamily\footnotesize,
    emph={self},
    emphstyle=\color{black}\underline,
    breakatwhitespace=false,         
    breaklines=true,                 
    captionpos=b,                    
    keepspaces=true,                 
    numbers=left,                    
    numbersep=5pt,                  
    showspaces=false,                
    showstringspaces=false,
    showtabs=false,                  
    tabsize=4
}

\usepackage{csquotes} 
\usepackage[backend=biber,citestyle=ieee,style=numeric-comp,sorting=none]{biblatex}

\definecolor{maroon}{RGB}{128, 0, 0}
\hypersetup{colorlinks=true,linkbordercolor=red,linkcolor=maroon}

\begin{document}

\title{Multimodal Representation for Neural Code Search
}

\author{
\IEEEauthorblockN{Jian~Gu}
\IEEEauthorblockA{\textit{University of Zurich}}
gu@ifi.uzh.ch
\and
\IEEEauthorblockN{Zimin~Chen}
\IEEEauthorblockA{\textit{KTH Royal Institute of Technology}}
zimin@kth.se
\and
\IEEEauthorblockN{Martin~Monperrus}
\IEEEauthorblockA{\textit{KTH Royal Institute of Technology}}
martin.monperrus@csc.kth.se
}

\maketitle
\vspace{-10em}

\begin{abstract}
    Semantic code search is about finding semantically relevant code snippets for a given natural language query. In the state-of-the-art approaches, the semantic similarity between code and query is quantified as the distance of their representation in the shared vector space. In this paper, to improve the vector space, we introduce tree-serialization methods on a simplified form of AST and build the multimodal representation for the code data. We conduct extensive experiments using a single corpus that is large-scale and multi-language: CodeSearchNet. Our results show that both our tree-serialized representations and multimodal learning model improve the performance of code search. Last, we define intuitive quantification metrics oriented to the completeness of semantic and syntactic information of the code data, to help understand the experimental findings.
\end{abstract}


\section{Introduction}
\label{sec:introduction}

\IEEEPARstart{I}{n} modern society, software systems are indispensable and have already been everywhere. It represents a wide range of applications, such as code search, which is the task to search existing code snippets. When developing or maintaining software, people tend to reuse existing scaffolds or learn from actual usage examples instead of wasting time reinventing the wheel. On one hand, code search could assist programmers in their daily work. On the other hand, it strengthens the infrastructures in open source communities, such as GitHub \footnote{https://github.com/}. It is beneficial to have an efficient and effective way for code search, comparing with the general-purpose search using modern search engines \cite{Rahman2018EvaluatingHD}.

The task of semantic code search is to retrieve the most semantically relevant code snippets for the given natural language queries. The input data is a query, and output data is an ordered list of code snippets. The code snippets ranked higher should be more similar to the query with regards to the text meaning. In state-of-the-art approaches, the code and query data are represented as vectors of the same length, so the computation of their semantic similarity is converted as the vector distance measurement.

We propose utilizing multimodal code representations to improve the semantic code search. Modality refers to the channels how information exists, and multimodal means the multiple modalities, namely information of multiple types. We introduce tree-serialized representations, to study whether they are more informative than the token representation and whether using them as the additional input is beneficial or not. Tree-serialized representations are generated by parsing code data into the tree structure and then serializing the tree into sequences. We utilize multimodal learning to reveal the power of multimodal representations and evaluate our approach by comparing with the given baselines on the CodeSearchNet corpus \cite{Husain2019CodeSearchNetCE}. Besides, we define two intuitive metrics, namely link coverage and node coverage, to quantify the completeness of the syntactic and semantic information conveyed by various code representations separately.

In our experiments, the SelfAtt model is the strongest model considering the context information. Among tree-serialized representations, traversal-based representations always perform better than sampling-based representations. Based on our results, tree-traversal representation improves the MRR scores by at most 16.88\%. More results suggest that our multimodal approach surpasses baselines by from 5.23\% to 15.70\%. Compared with corresponding unimodal representations, multimodal representations bring improvements for at most 17.82\%. Furthermore, we recommend adopting tree representations as well as multimodal learning for code search.

The contributions of this paper are as follows:

\begin{enumerate}
  \item We propose the Simplified Semantic Tree, a specifically designed form of Abstract Syntax Tree for enriching the semantic representation of the program. We systemically define and assess several serialization methods based on this novel tree structure.
  
  \item We demonstrate that multimodal learning is effective for the code search task. We compare the representation capability of both unimodal and multimodal code search using our tree-serialized representations. To the best of our knowledge, it is the first time that tree-serialized sequences are used as the modality for code search.

  \item We define two simple but intuitive quantification metrics, the link coverage and the node coverage, to measure the completeness of semantic and syntactic information of various representation forms. They perform reliably in revealing the effectiveness of modalities for code search.
\end{enumerate}

The rest of this paper is structured as follows. In \autoref{sec:background}, the background information is introduced. In \autoref{sec:approach}, our approach on multimodal representation is proposed. In \autoref{sec:methodology}, the experimental methodology is described. In \autoref{sec:results}, the experimental results are shown and explained. In \autoref{sec:discussion}, threats to validity are discussed. In \autoref{sec:related}, related work and research context are reviewed. In \autoref{sec:conclusion}, the whole work is summarized.

\section{Background}
\label{sec:background}

\subsection{Code Search}

The canonical task of code search is about finding the most relevant code snippet for the given natural language query. A code search engine can be made with information-retrieval techniques and neural techniques \cite{Yan2020AreTC}. In this paper, we focus on the latter.

As illustrated in \autoref{lst:demo}, a code-query pair is a piece of natural language query and the corresponding source code. The query could be short documentation of the target code snippet, like ``send birthday messages to members''. Throughout this work, we call the source code tokens ``code sequence'' and the query tokens ``query sequence''. The code and query sequences are used to train a neural code search model.

\begin{lstlisting}[language=Python,label={lst:demo},caption=A Code Snippet in Python]
def birthday_marketing(self):
    """send birthday messages to members"""
    today = datetime.date.today()
    for member in self.members:
        birthday = member.birthday
        if self.anniversary(today, birthday):
            member.SMS()
\end{lstlisting}

\subsection{Siamese Networks for Code Search}
\label{sec:intro-siamese}

\begin{figure}[!htb]
  \centering
  \includegraphics[width=\linewidth]{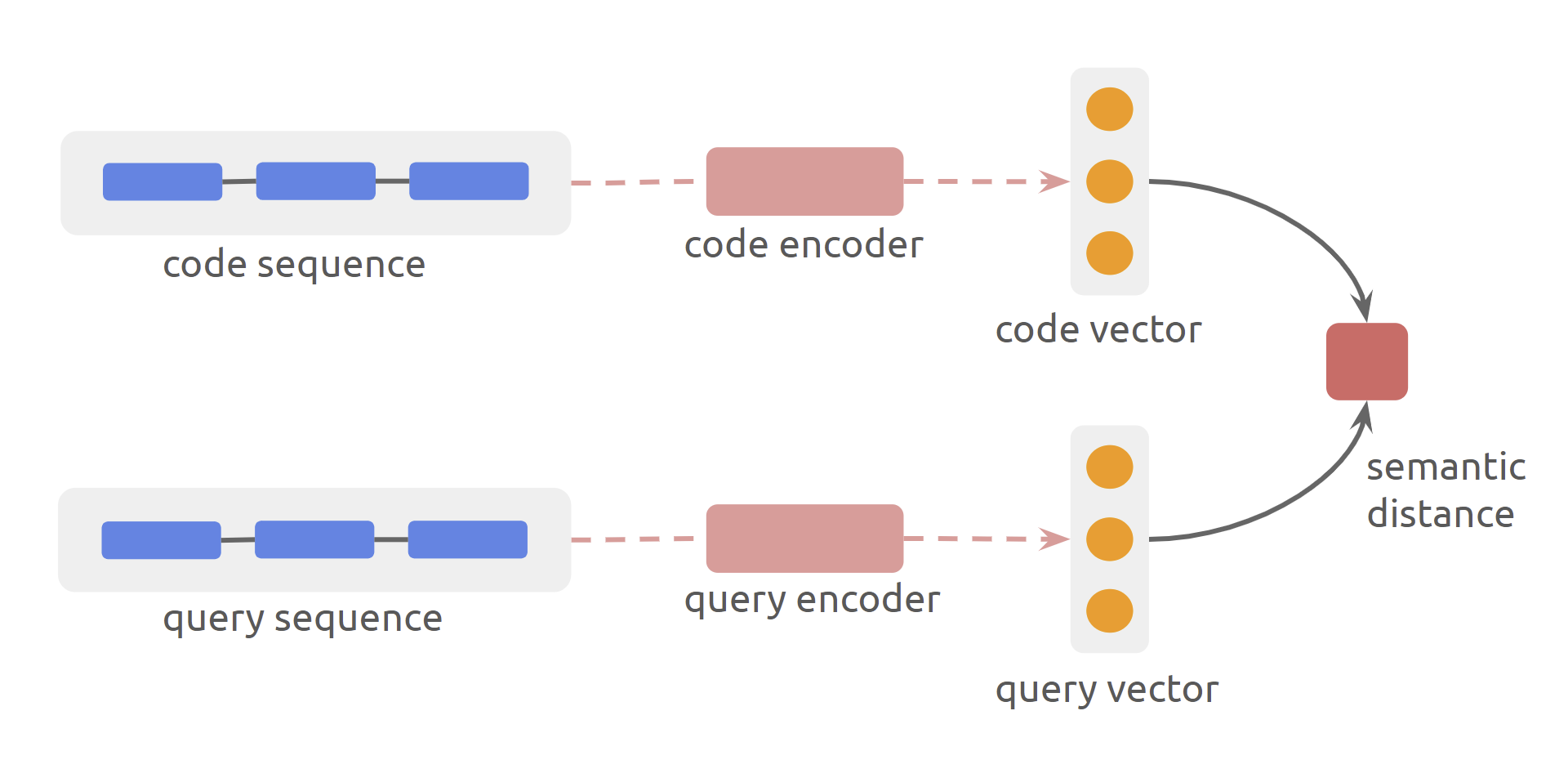}
  \caption{Pseudo-Siamese Network for Code Search \cite{Husain2019CodeSearchNetCE}}
  \label{fig:3-architecture}
\end{figure}

A siamese network is an artificial neural network to measure the similarity between two inputs of the same type using the same encoder \cite{Zagoruyko2015LearningTC}. The pseudo-siamese network is more flexible because it is intended to measure the similarity of different data types with different encoders \cite{Hughes2018IdentifyingCP}. The model architecture for code search follows the practice of utilizing the pseudo-siamese network \cite{Gu2018DeepCS}, as illustrated in \autoref{fig:3-architecture}. In the architecture, the code and query sequences, are respectively fed into corresponding encoders to be converted into vectors. The training objective is to minimize the distance between relevant code and query vectors. The code search model uses the cosine distance between vectors to measure the similarity. Once trained, the most semantically relevant code snippets to a given query, are those whose vectors are closest to the query vector measured by cosine distance.

The learning objective is to make sure that semantically similar vectors are as close as possible. The \emph{triplet loss} is the objective function used in baselines and our model. It optimizes the query vector to be close to the corresponding code vector, but to be far away from other code vectors.
In the training process, each code-query pair $(c_i, q_i)$ and corresponding distractor code snippets $c_j$ are fed into the code encoder $E_c$ and query encoder $E_q$. The training objective is to minimize the loss shown below:

\begin{equation}\mathrm{Loss}=-\frac{1}{N} \sum_{i} \log \left(\frac{\exp \left(E_{c}\left(\mathbf{c}_{i}\right)^{\top} E_{q}\left(\mathbf{q}_{i}\right)\right)}{\sum_{j} \exp \left(E_{c}\left(\mathbf{c}_{j}\right)^{\top} E_{q}\left(\mathbf{q}_{i}\right)\right)}\right)\end{equation}

The triplet loss function is aimed to maximize the inner product of code $c_i$ and query $q_i$ encodings of the pair, while minimize the inner product between target code snippet $c_i$ and its distractor code snippets $c_j$ ${i}\neq{j}$ \cite{Husain2019CodeSearchNetCE}. $(c_i, q_i)$ and $(c_j, q_i)$ respectively denotes the positive and negative samples.

There is a wide diversity of possible encoders for siamese networks for code search. Dedicated to sequential data, typical encoders include:
\begin{itemize}
  \item \emph{NBoW:} Neural Bag-of-Words (NBoW) simply computes the weighted average of all word embeddings to get the sentence embedding as the whole semantic representation \cite{Kalchbrenner2014ACN,Iyyer2015DeepUC}.
  \item \emph{1D-CNN:} Convolutional Neural Network (CNN) \cite{Kim2014ConvolutionalNN} uses a convolution operation to analyze context information in receptive fields of different sizes. 1D-CNN refers to the model for 1-dimension sequential data.
  \item \emph{Bi-RNN:} Recurrent Neural Network (RNN) \cite{Elman1990FindingSI} uses hidden layers in the form of temporal sequence to capture dependencies. Bi-RNN concatenates embeddings of both forward and backward directions. 
  \item \emph{SelfAtt:} The transformer-based model uses a self-attention mechanism and the positional embedding way of BERT to learn from the context information \cite{Vaswani2017AttentionIA,Devlin2019BERTPO}.
\end{itemize}

\subsection{Multimodal Learning}

Modality refers to the way how some type of information exists. For example, to identify shepherd dogs from sheep, we could fully utilize the data of various modalities, such as colors, sounds, and features of movement patterns.

Multimodal learning aims to build models that are capable to process and associate data of multiple modalities \cite{Baltruaitis2019MultimodalML}. It is based on the fact that data semantics can be captured in different ways. The representation resulting from data of multiple modalities is named as the \emph{multimodal representation}. Because the multimodal learning model learns features considering the information from various modalities, it usually performs better than the unimodal learning model which only studies data of unique modality \cite{Ngiam2011MultimodalDL}.

\section{A Novel Multimodal Representation Approach for Code Search}
\label{sec:approach}

In this section, we present a novel approach for neural code search. This approach follows the design of the pseudo-siamese network for code search.
Our core intuition is to build an encoder that fully utilizes information from multiple aspects of source code. In this paper, such an aspect is called a ``modality'', per the seminal work of multimodal learning with deep Boltzmann machines \cite{Srivastava2012MultimodalLW}.
\subsection{Overview}

\begin{figure}[!htb]
  \centering
  \includegraphics[width=0.5\linewidth]{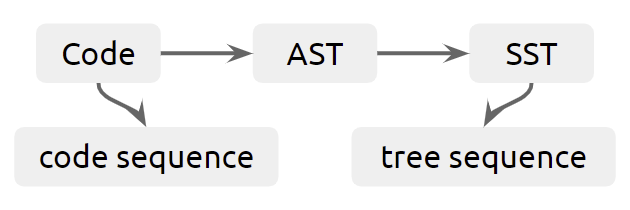}
  \caption{Workflow for Extracting Multiple Representations}
  \label{fig:3-workflow}
\end{figure}

Based on the raw code snippet of the particular code query, we extract code representations from multiple modalities. The workflow is as shown in \autoref{fig:3-workflow}.

The input data is the code snippet and the output data are the code sequence and tree sequence. First, we parse the code snippet into an \emph{Abstract Syntax Tree} (AST). To make the tree structure semantically better for code search, we transform the AST into a novel tree structure named \emph{Simplified Semantic Tree} (SST), which we newly introduce in this paper. Then, we extract a \emph{tree-serialized representation} from the SST, by sampling tree-paths \cite{Alon2018AGP,Kim2020CodePB} or traversing tree-structures \cite{Hu2018DeepCC,Chen2018TreetotreeNN}. Eventually, these tree-serialized representations separately complement the conventional \emph{token representation}, namely the sequence of source code tokens, in our multimodal learning approach.

\subsection{Simplified Semantic Tree}
We design one novel tree structure named SST for extracting tree-serialized representations of source code. SST is an approximation to simplify the tree structure of AST and highlights the semantic information of the code snippet. Compared with AST, SST removes unnecessary tree nodes and improves the label of tree nodes. Even though AST already has the tree structure and is capable of tree serialization, SST is more semantically informative and more general to various programming languages.

For any given AST, we conduct three operations to build the corresponding SST. The first operation is to prune tree nodes that are semantically meaningless for code search, such as type declarations like ``int'' and ``boolean'', modifier keywords like ``public'' and ``abstract'', functional keywords like ``async'' and ``await''. The complete list of removed nodes varies for different languages. The second operation is to replace labels of statement nodes and expression nodes with descriptive tags, like using ``loop'' for for-loop and while-loop statements and ``literal'' for exact string variables, the goal is to help the network grasp the general concept behind syntactically different nodes. The third operation is to unify the expression of semantically similar labels from different languages, like unifying ``function'', ``program'', ``define'', and ``module'' as ``module''. It is expected to promote some form of transfer learning across programming languages.

\begin{figure}[!htb]
  \centering
  \includegraphics[width=\linewidth]{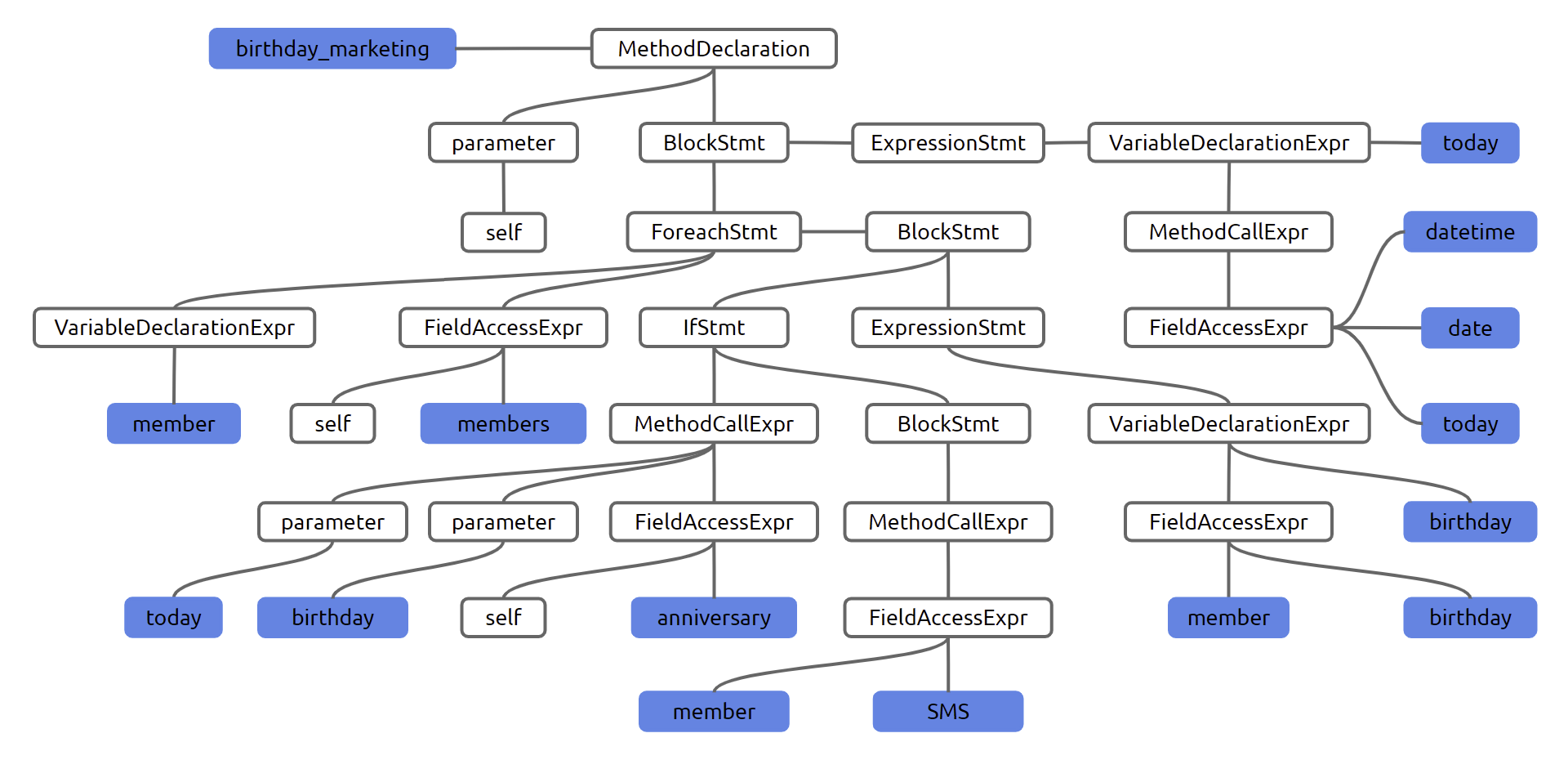}
  \caption{AST Diagram of \autoref{lst:demo}}
  \label{fig:4-demo-ast}
\end{figure}

To better understand SST, let us study the code snippet illustrated in \autoref{lst:demo}. This code iterates over all members and checks whether today is someone's birthday, if yes, it invokes the SMS function. These words in blue are identifiers. As illustrated in \autoref{fig:4-demo-ast}, the code snippet is parsed into AST. In the tree structure, most leaf nodes in the blue blocks are meaningful, while other leaf nodes seem semantically meaningless, such as the underlined ``self'', which is merely meant to explicitly represent the instance of the class. In contrast to leaf nodes, most non-terminals nodes correspond to keywords, punctuations, and indents.

\begin{figure}[!htb]
  \centering
  \includegraphics[width=\linewidth]{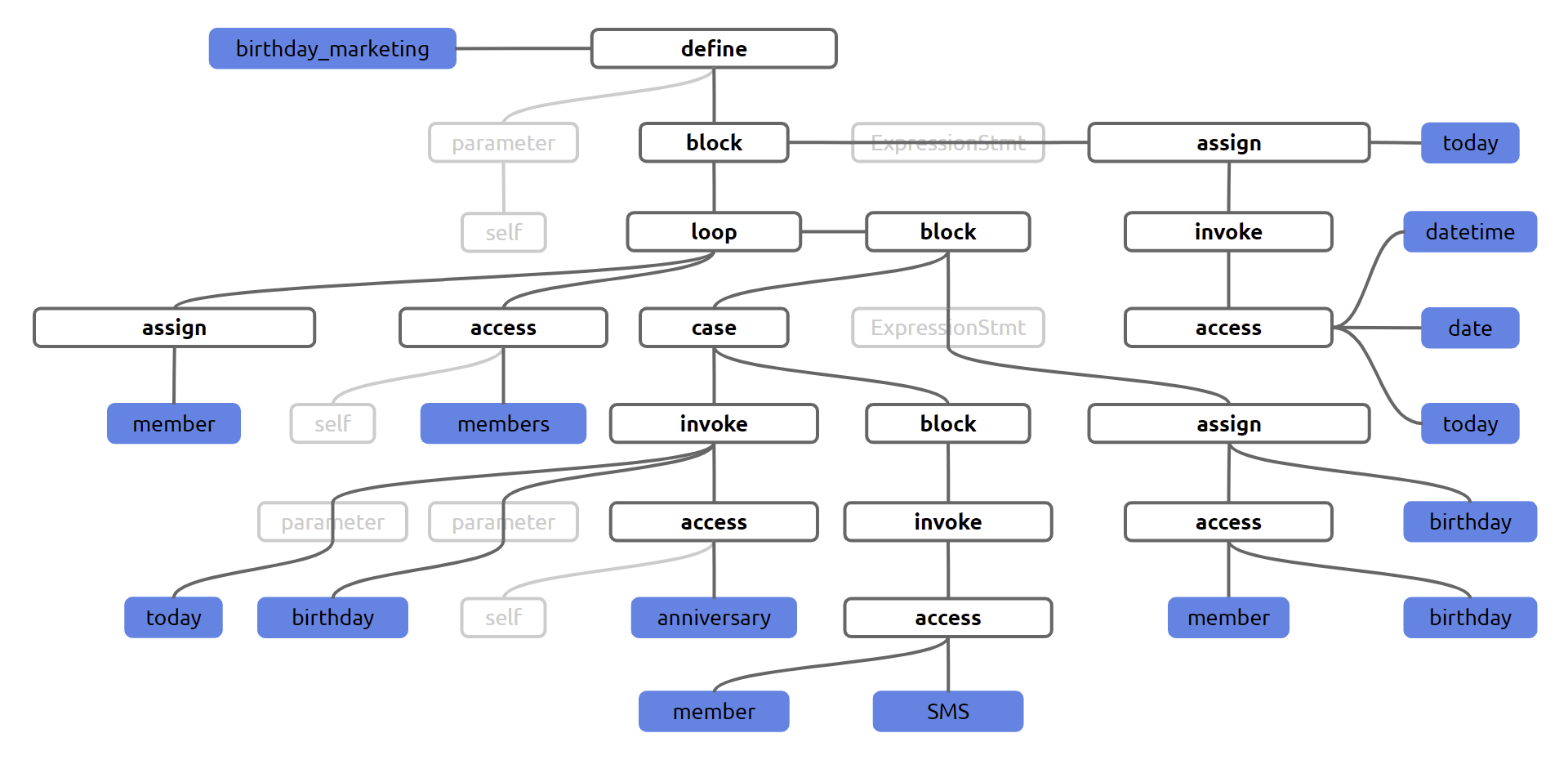}
  \caption{SST Diagram of \autoref{lst:demo}}
  \label{fig:4-demo-sst}
\end{figure}

Now turn to check the corresponding SST in \autoref{fig:4-demo-sst}. In the tree structure, some leaf nodes are removed (grayed out in the figure).
The non-terminal nodes are replaced with unified semantic expressions, that are shared among languages. For instance, the labels containing ``-Stmt'' or ``-Expr'' postfix are replaced by the shorter, cross-language version (eg. ``ForEachStmt'' into ``loop''). If observe the differences between AST and SST, the removed nodes are in gray and the simplified labels are in bold.

\subsection{Tree Serialization}
\label{subsec:serialization}

Next, we serialize the SST, per the technique of sampling and traversal. The motivation is to extract a linear sequence from the tree structure. These sequences are sequential data that are more applicable to the typical encoders above mentioned.
There are two options to serialize SST to a token sequence. One way is to extract tree paths from the tree structure \cite{Alon2018AGP,Kim2020CodePB}, namely the connections between tree nodes, and then filter and sample over the collected tree paths. The other way is to serialize the whole structure via tree traversal \cite{Hu2018DeepCC,Chen2018TreetotreeNN}. They are respectively called as \emph{sampling-based representation} and \emph{traversal-based representation}.
In this paper, we investigate two sampling-based representations and two traversal representations.

\begin{figure}[!htb]
  \centering
  \includegraphics[width=\linewidth]{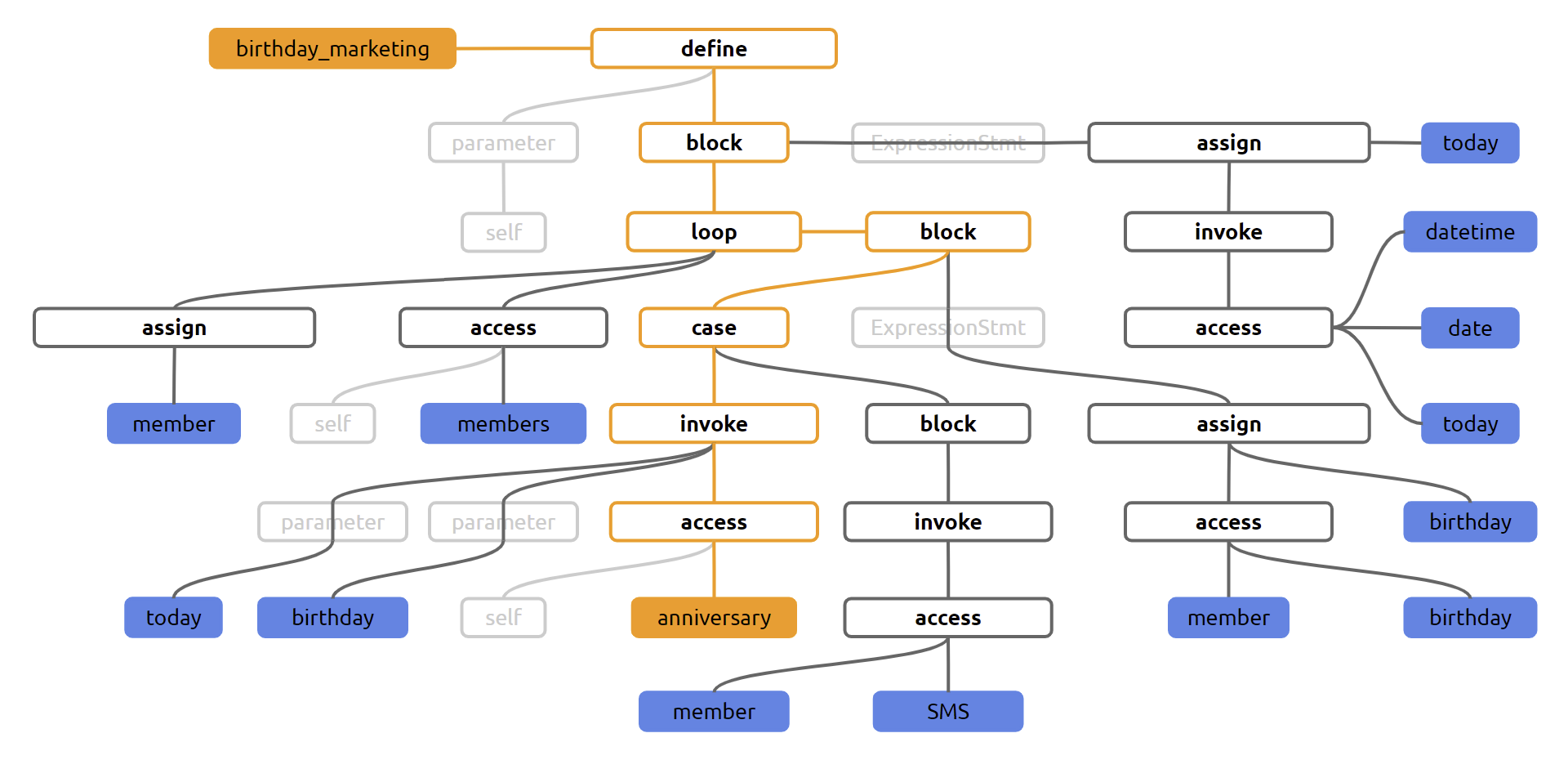}
  \caption{RootPath Schematic Diagram}
  \label{fig:4-demo-rootpath}
\end{figure}

\textbf{RootPath} \cite{Alon2018AGP} samples the paths of non-terminal nodes starting from a single leaf node to the root node. As shown in \autoref{fig:4-demo-rootpath}, it is such one rootpath connecting the leaf node whose label is ``anniversary'' and the root node. All nodes at the tree path are in the orange blocks.

\begin{figure}[!htb]
  \centering
  \includegraphics[width=\linewidth]{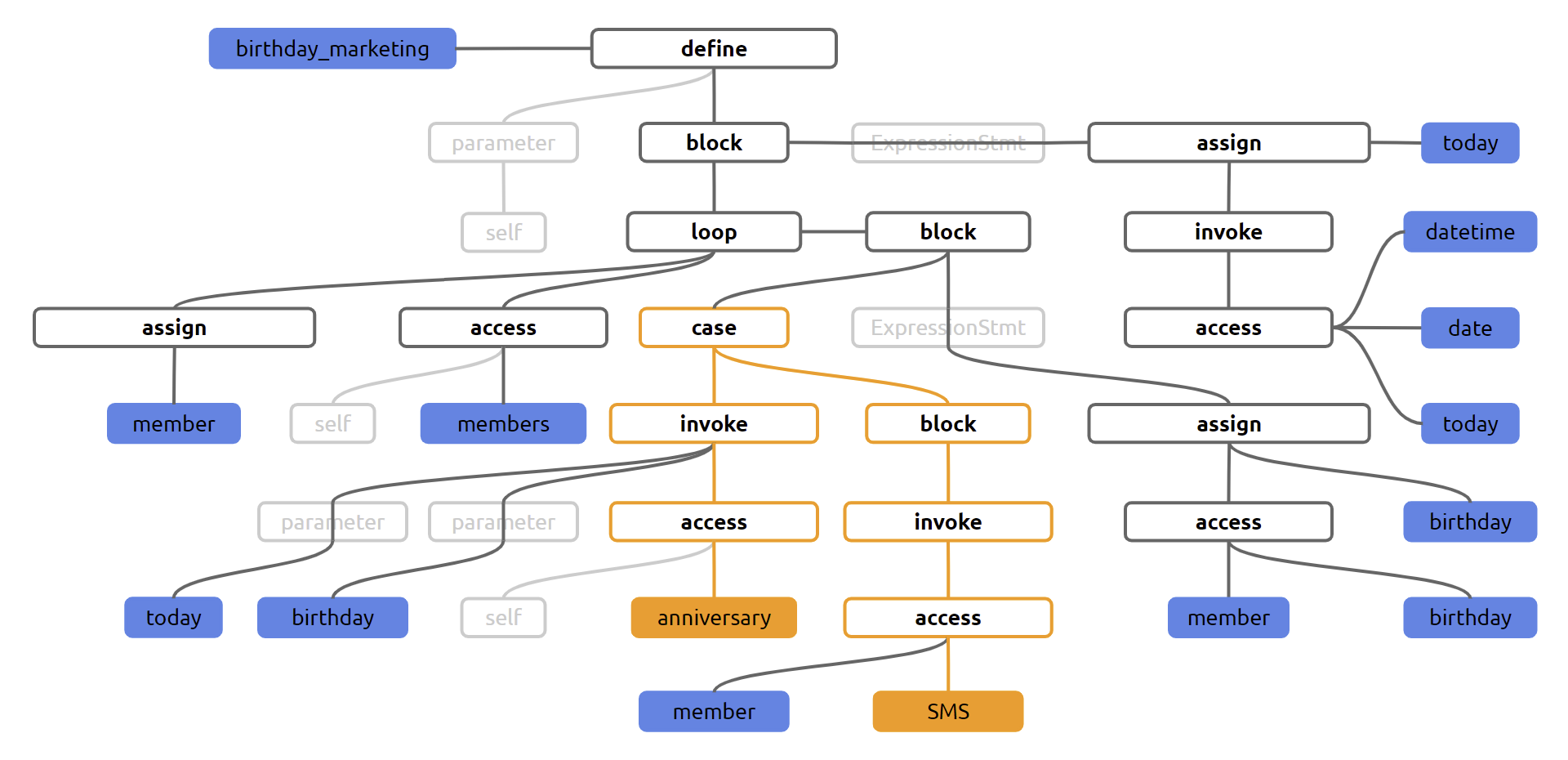}
  \caption{LeafPath Schematic Diagram}
  \label{fig:4-demo-leafpath}
\end{figure}

\textbf{LeafPath} \cite{Kim2020CodePB} samples the paths of non-terminal nodes between two arbitrary leaf nodes. As shown in \autoref{fig:4-demo-leafpath}, it is such one leafpath connecting the leaf node whose label is ``anniversary'' and another leaf node whose label is ``SMS''. All nodes at the tree path are in the orange blocks.

\textbf{Structure-Based Traversal} (SBT) \cite{Hu2018DeepCC} is a traversal-based representation. SBT Representation is obtained via top-down recursive tree traversal. The detailed procedure is as follows: 1) from the root node, we first use a pair of brackets to represent the tree structure and put the root node itself behind the right bracket. 2) traverse the subtrees of the root node and put all root nodes of subtrees into the brackets. 3) traverse each subtree recursively until all nodes are traversed and the final sequence is obtained.

\begin{figure}[!htb]
  \centering
  \includegraphics[width=\linewidth]{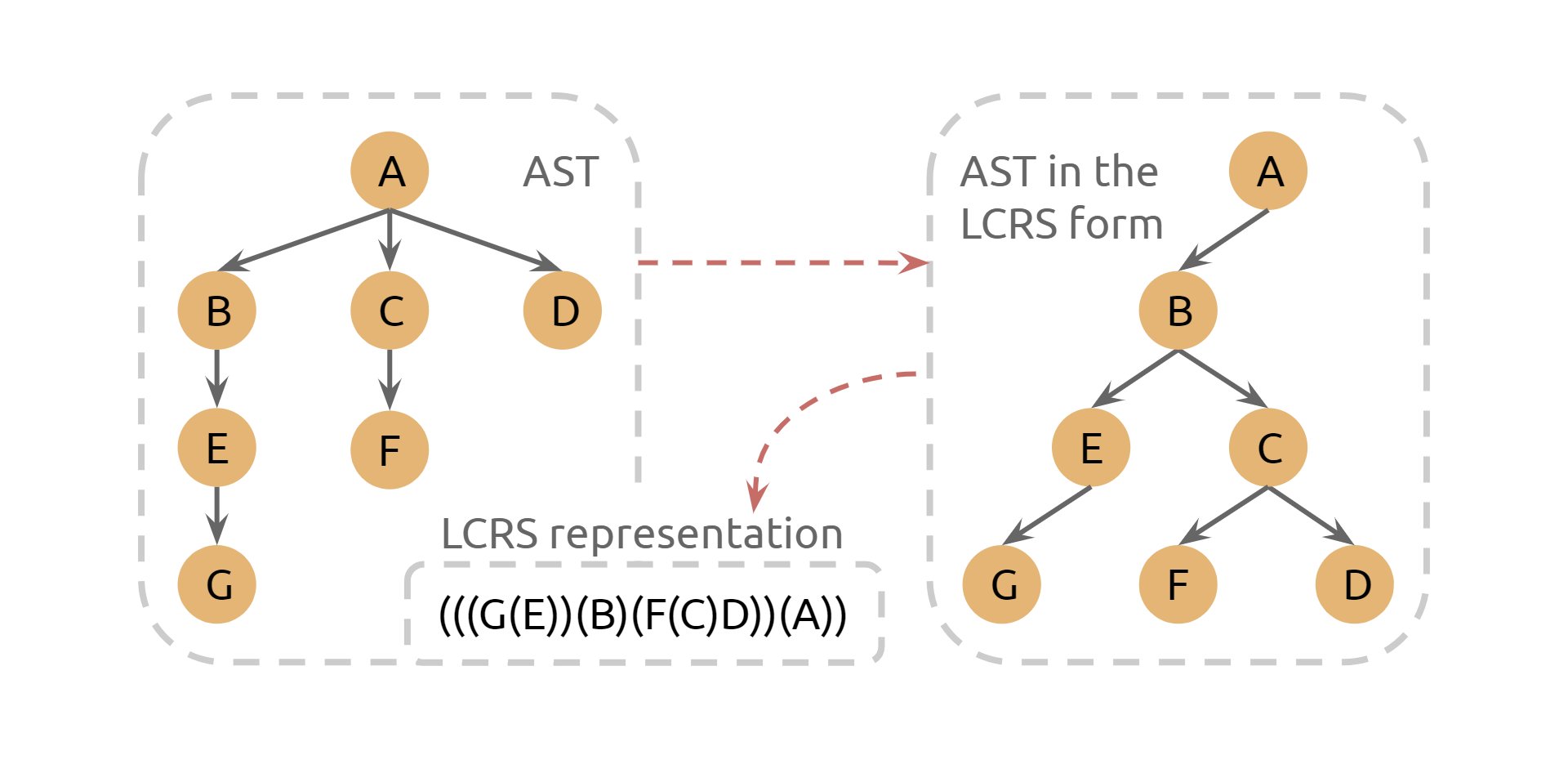}
  \caption{LCRS Representation}
  \label{fig:4-schema-lcrs}
\end{figure}

\textbf{Left-Child Right-Sibling} (LCRS) \cite{Chen2018TreetotreeNN} is a traversal-based representation. The idea is to transform the general tree into a binary tree in the Left-Child Right-Sibling form, as illustrated in \autoref{fig:4-schema-lcrs}. Then LCRS representation is obtained via in-order tree traversal. For each subtree, we use brackets to separate the parent node, left child tree, and right child tree.

\begin{figure}[!htb]
  \centering
  \includegraphics[width=\linewidth]{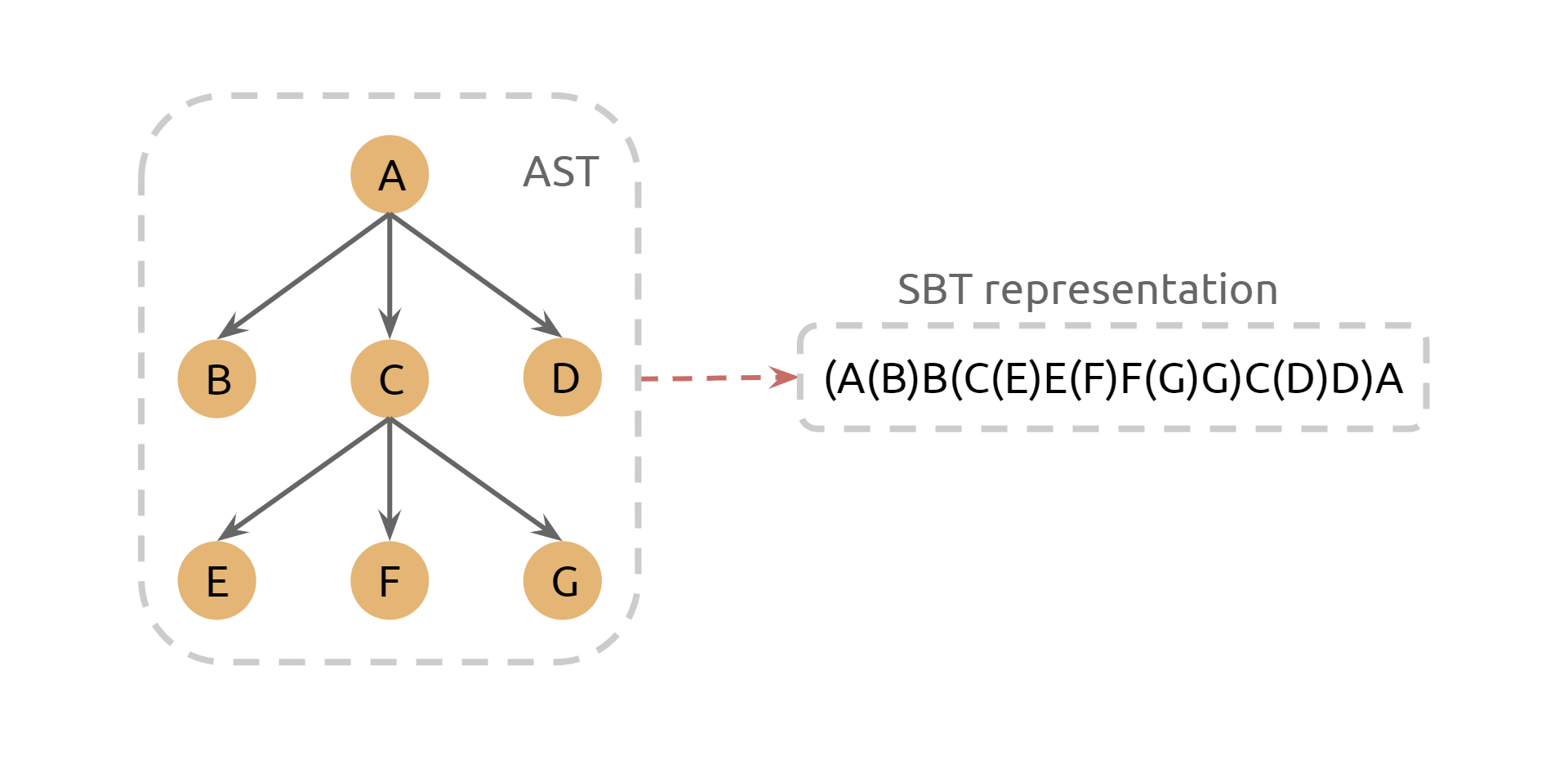}
  \caption{SBT Representation}
  \label{fig:4-schema-sbt}
\end{figure}

To generate these tree-serialized representations, we parse raw code snippets to build SST, then extract rootpaths starting from each leaf node whose label is an identifier to the root node. For better semantic expressions, we ignore leaf nodes whose labels are single-character identifiers, like ``t'' or ``x'', unless there are no enough rootpaths for each code snippet. Once we collect all rootpaths, we combine them into pairs randomly to generate leafpaths. Similarly, we give the priority to leafpaths whose corresponding leaf nodes are with multi-character identifiers, like ``cost'' or ``ratio'', as labels, because they are seen as most semantically informative. In contrast, we only need to implement functions for structure-based traversal and in-order traversal as well as the tree transformation algorithm to generate the SBT and LCRS representations, without any extra processing work.

The RootPath and LeafPath representations are controlled by a number called sampled paths $N$. If $M$ is the number of nodes in the tree, there are $M$ unique rootpath sequences and $M*(M-1)/2$ unique leafpath sequences.
For the LeafPath representation, we discard low-quality leafpath sequences, whose sequence length is larger than the suggested $length_{threshold}=8$ or the tree height difference of two sides is smaller than the suggested $width_{threshold}=2$ \cite{Alon2018code2vecLD,Alon2018code2seqGS}. Meanwhile, the default parameter for the number of sampled tree paths is $N=20$. A higher value promises better results but requires more computing power, so here is a trade-off.

\subsection{Multimodal Learning}

In our approach, we extract two modalities from source code, one modality from the natural language representation, and adopt multimodal learning on top of them. The two modalities are plain and tree-serialized representations of code snippets. In the following, the modality of token representation is called the \emph{code sequence} and represents tokens. The tree-serialized representation of SST is called the \emph{tree sequence}.

\begin{figure}[!htb]
  \centering
  \includegraphics[width=\linewidth]{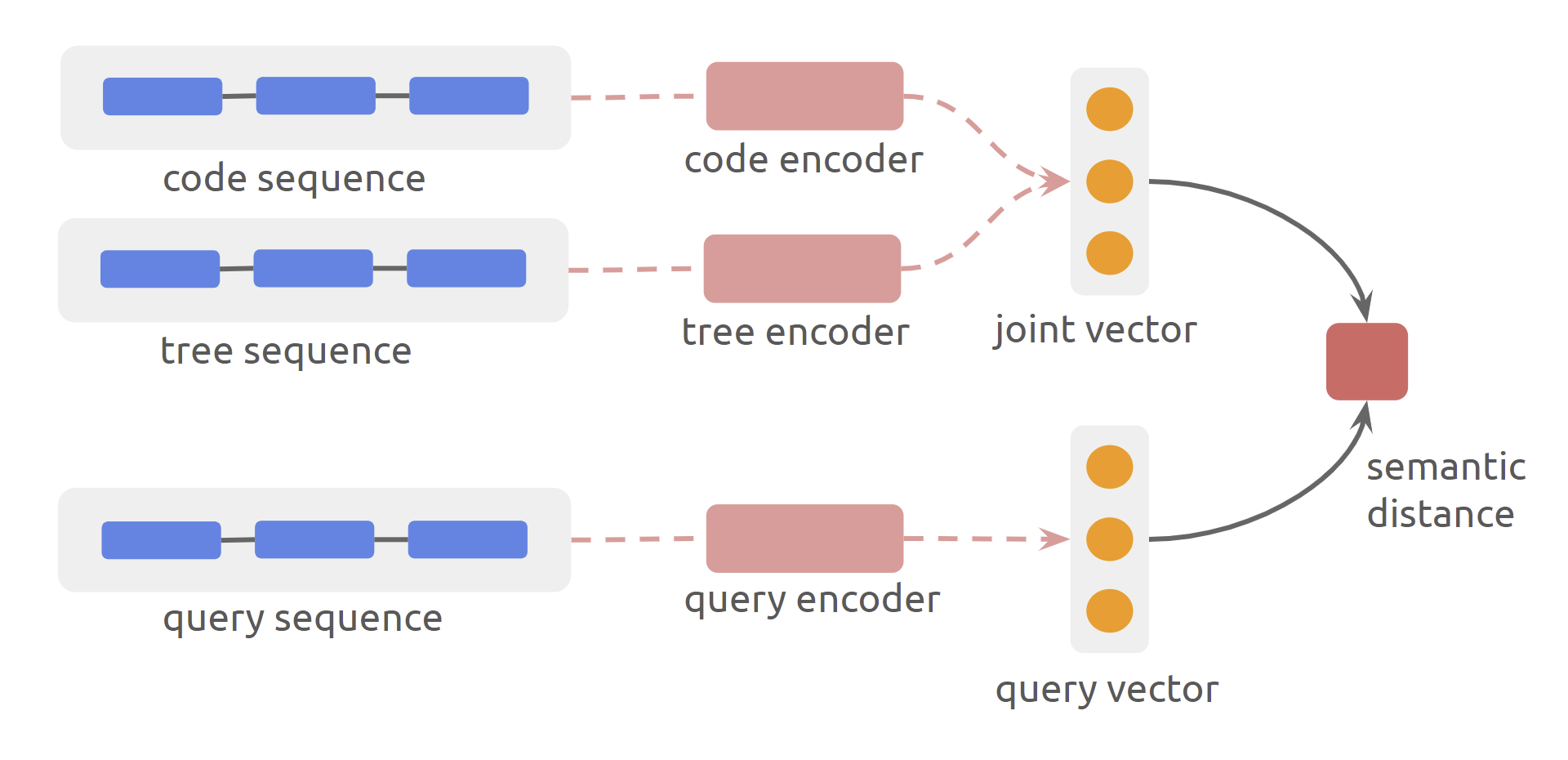}
  \caption{Multimodal Learning Schematic Diagram}
  \label{fig:4-modality}
\end{figure}

The architecture of our multimodal learning model is a pseudo-siamese architecture, as shown in \autoref{fig:4-modality}. We adopt three SelfAtt models as the encoders in our model, considering that the SelfAtt model performs very well in various natural language tasks. The code and tree encoders are working in parallel for the code representation, and the query encoder is for the query representation. We feed the code encoder with code sequence and tree encoder with tree sequence, and these two encoders convert their respective sequence data to vectors, to a token vector and tree vector. The query encoder receives a query sequence as input and computes the query vector as output. More specifically, SBT or LCRS representation is used as the input data for the tree encoder, while other tree-serialized representations are applicable as well.

In our multimodal learning model, all three modality vectors are of the same length. We combine the code and tree vectors by summing them. This computes a joint vector which is the multimodal representation for the source code. Then the joint vector is trained together with the query vector, to ensure that semantically similar code vectors are close to the query vector in the shared vector space. As usual, we compute the cosine distance to quantify the semantic similarity.

\section{Experimental Methodology}
\label{sec:methodology}

The work investigates the following research questions:

\newcommand\rqaaa{What is the performance of siamese networks for code search when using different encoders?}
\newcommand\rqbbb{What is the relative performance of the considered unimodal tree-serialized representations?}
\newcommand\rqccc{What is the effectiveness of multimodal representations for code search?}

\begin{itemize}
  \item \textbf{RQ1:} \rqaaa
  \item \textbf{RQ2:} \rqbbb
  \item \textbf{RQ3:} \rqccc
\end{itemize}

\subsection{Corpus}

In our experiment, we use the CodeSearchNet corpus as the experimental data \cite{Husain2019CodeSearchNetCE}. The corpus contains over two million code-query pairs in six popular programming languages, namely Go, Java, JavaScript, PHP, Python, and Ruby.

As shown in Table \ref{tab:0-corpus}, the corpus is split into datasets with the 80-10-10 proportion. The first column presents datasets. The second to the seventh columns present the statistics of each programming language. The last column is for the total statistics values for all languages.

\begin{table}[thb]
\centering
\caption{Statistics of Code-Query Pairs in the Corpus}
~\\
\label{tab:0-corpus}
\resizebox{\linewidth}{!}{
\begin{tabular}{lccccccc}
\toprule
Dataset & Go & Java & JavaScript & PHP & Python & Ruby & Total \\
\midrule
Train & 317,832 & 454,451 & 123,889 & 523,712 & 412,178 & 48,791 & 1,880,853 \\
Valid & 14,242 & 15,328 & 8,253 & 26,015 & 23,107 & 2,209 & 89,154 \\
Test & 14,291 & 26,909 & 6,483 & 28,391 & 22,176 & 2,279 & 100,529 \\
All & 346,365 & 496,688 & 138,625 & 578,118 & 457,461 & 53,279 & 2,070,536 \\
\bottomrule
\end{tabular}
}
\end{table}

\subsection{Performance Metrics}

The challenge uses the \emph{mean reciprocal rank} (MRR) and the \emph{normalized discounted cumulative gain} (NDCG) score, as the performance metrics. The MRR score is used to measure models by competitors before their submissions, then the NDCG score is computed as the further criteria in the competition. The higher MRR or NDCG scores indicate better performance. The computation of the former is on the testing set, and the latter is on a small-sized artificial evaluation set.

The MRR score quantifies the ranking of the target code snippet to the given query, and it only cares about where the most relevant result is ranked. The most relevant code snippet should be ranked the highest, the lower its ranking position, the lower the MRR score. When computing MRR scores in the testing set, for each code-query pair, 999 code snippets from other pairs in the same batch play the role of distractors. The average value of all batches is the final MRR score.



The NDCG score quantifies the similarity between the rankings of candidate code snippets and the most ideal rankings, and it cares about the whole ranking order between all candidate results. The most ideal case is when more relevant code snippets are always ranked before those less relevant ones. The computation of NDCG scores on the extra evaluation set was specially made for the challenge. The evaluation set consists of 99 natural language queries, and 10 candidate code snippets for each query. The NDCG score is computed on the whole evaluation set.

\subsection{Information Completeness Metrics}
\label{subsec:metric}

To intuitively measure the information completeness of code representations, we define \emph{link coverage} and \emph{node coverage} as the quantification metrics. The link coverage is to measure the completeness of syntactic information, and the node coverage is to measure the completeness of the semantic information. The coverage ratio is that of tree components, e.g. tree links or tree nodes, to be utilized among all components. Therefore, higher coverage indicates better informativeness or more complete information content.

The link coverage is defined as follows that how many tree links between each pair of nearest tree nodes are taken from all the tree links, so the syntactic information is expressed as to which extent the tree links between each pair of the nearest nodes are covered by tree sequences.

The node coverage is defined as follows that how many tree nodes are taken from all the tree nodes, so the semantic information is expressed as to which extent the unique labels of tree nodes are contained by tree sequences. We simply count tree nodes sharing the same label as one unique tree node.

Taking the SST diagram shown in \autoref{fig:4-demo-sst}, the rootpath and leafpath demonstrated in \autoref{fig:4-demo-rootpath} and \autoref{fig:4-demo-leafpath} as an example. For the node coverage, there are 31 tree links totally because we directly count the links between the nearest tree nodes. Follow the same idea, the rootpath and leafpath separately cover 8 and 7 links, so their link coverages are 25.81\% and 22.58\%. For the node coverage, there are 32 tree nodes in total but sharing 16 unique labels, so the SST only has 16 unique tree nodes. Similarly, the rootpath and leafpath separately contain 8 and 6 unique nodes, so their node coverages are 50.00\% and 37.50\%.

The formulas for the link coverage and node coverage are defined as follows:

\begin{equation}\mathrm{LinkCoverage}_{\mathrm{C}}=\frac{\left|links_{C}\right|}{\left|links_{T}\right|}=\frac{\left|\cup_{i}^{n}{links_{S_i}}\right|}{\left|links_{T}\right|}\end{equation}

\begin{equation}\mathrm{NodeCoverage}_{\mathrm{C}}=\frac{\left|nodes_{C}\right|}{\left|nodes_{T}\right|}=\frac{\left|\cup_{i}^{n}{nodes_{S_i}}\right|}{\left|nodes_{T}\right|}\end{equation}

where $C$ represents the collection of sequences $S_{i}$, and $T$ is the tree where sequences $S_{i}$ are generated. Besides, $links_{O}$ and $nodes_{O}$ represent the tree links and tree nodes from a given object $O$, such as a sequence, a collection of sequences, or a tree, e.g. AST or SST.

According to the definitions, for a given code snippet, the link coverage and node coverage of token representation are respectively 0\% and 100\%. For tree-serialized representation, if sampling-based, like RootPath or LeafPath, these ratios are between 0\% and 100\%, if traversal-based, for SBT, both ratios are 100\%, and for LCRS, the node coverage is 100\%. Besides, if compute over the combination of the introduced representations. The definitions are still valid because different representations are from the same tree structure.

Even though the link coverage and node coverage could indicate how much syntactic and semantic information the input data may contain, it does not mean that more information always brings better results. Besides, considering the trade-off between the information completeness of the input data and the complexity of the feature extractor, it is not feasible to directly make a whole tree as the input data without introducing more work into model designing and training.

\subsection{Methodology for RQ1}
\label{subsec:mrq1}

\emph{\rqaaa}

The CodeSearchNet Challenge provides four baselines. We independently measure their performance to find the strongest baseline model.

We run experiments over the whole CodeSearchNet corpus, and we compute the MRR score over the testing set and the NDCG score over the evaluation set. For each baseline model, it is trained and tested over the data of each programming language. If one model has a high MRR score, it means it can learn well, when the samples are drawn from the same distribution. If one model has a high NDCG score, then it indicates a strong generalization, because the characteristics of the training set and the evaluation set are not alike.

\subsection{Methodology for RQ2}
\label{subsec:mrq2}

\emph{\rqbbb}

Our goal is to compare the performance of tree-serialized representations introduced in \autoref{subsec:serialization} against the token representation. In all experiments, we feed the SelfAtt model with the tree sequence, not the code sequence. There are no changes to other settings compared to RQ1.

We name the experimental model of using the token representation as Uni-Code, where ``Uni-'' refers to using a unimodal setup. Uni-RootPath, Uni-LeafPath, Uni-SBT, or Uni-LCRS refers to the unimodal learning with RootPath, LeafPath, SBT, or LCRS representation respectively.

To make our experiments more representative, we focus on the Python and Ruby corpora, considering the contrastive dataset size and language features. The performance of models is measured by the MRR score, because it is more reliable than NDCG, considering the NDCG score is computed on a small evaluation set and its numerical values tend to be rather low.

\subsection{Methodology for RQ3}
\label{subsec:mrq3}

\emph{\rqccc}

The motivation for introducing tree-serialized representations is to complement the token representation because we are interested in studying the effectiveness of multimodal representations, namely the combination of token representation and tree-serialized representation. In this RQ, we measure the performance of our multimodal learning approach. There are three encoders in our multimodal learning model, one query encoder is for query sequence, then the code and tree encoders are respectively for the code and tree sequences. We trained four models with different tree serialization techniques while keeping other settings the same. Otherwise stated, the corpora and metrics are the same as for RQ2.

Following the naming way of Uni-RootPath, Uni-LeafPath, Uni-SBT, or Uni-LCRS, we name the experimental models of using the multimodal representation containing RootPath, LeafPath, SBT, or LCRS representation as Multi-RootPath, Multi-LeafPath, Multi-SBT, Multi-LCRS.

\subsection{Implementation}

Our implementations are written in Python, depending on the TensorFlow \cite{Abadi2016TensorFlowAS} framework. We adopt tree-sitter \footnote{https://github.com/tree-sitter/tree-sitter} for parsing code. For RQ1, we use the official implementation of baselines supplied by the CodeSearchNet challenge directly. For RQ2 and RQ3, we introduce some optimization operations to make baselines stronger, such as model tuning and indexing strategy refining. In all conducted experiments, the hyper-parameters remain unchanged, the batch size is 1000 and the embedding size is 128.

The experiments are conducted on the Ubuntu Linux with 56 GiB memory, powered by one piece of NVIDIA Tesla K80 GPU and six pieces of Intel Xeon E5-2690 CPU. The replication repository is available online \footnote{https://github.com/jianguda/mrncs}, which contains implementations, results, and the guidance for reproduction.

\section{Experimental Results}
\label{sec:results}

To study the mentioned research questions, we conduct experiments and compare results. The accuracy of models is reflected by the MRR and NDCG scores. In the following comparisons, the best results are highlighted in bold.

\subsection{Results of RQ1}

In this section, we compare the performance of different unimodal encoders for neural code search, per the protocols described in \autoref{sec:methodology}.

Table \ref{tab:1-mrr} and \ref{tab:1-ndcg} give the MRR score and NDCG score of the considered unimodal encoders respectively. The rows are for encoders and the columns are for the languages of the benchmark. For example, the MRR score and NDCG score of the NBow encoder on Go are respectively 0.6881 and 0.1165.

\begin{table}[thb]
\centering
\caption{RQ1: MRR Scores of Baseline Models}
~\\
\label{tab:1-mrr}
\resizebox{\linewidth}{!}{
\begin{tabular}{lccccccc}
\toprule
Encoder & Go & Java & JavaScript & PHP & Python & Ruby & Avg. \\
\midrule
NBoW & 0.6681 & $\mathbf{0.5867}$ & $\mathbf{0.4268}$ & 0.5679 & 0.6432 & 0.3210 & 0.5356 \\
1D-CNN & 0.7044 & 0.5302 & 0.2298 & 0.5429 & 0.5383 & 0.1165 & 0.4437 \\
Bi-RNN & 0.7082 & 0.5808 & 0.3685 & 0.6012 & 0.6432 & 0.2226 & 0.5208 \\
SelfAtt & $\mathbf{0.7257}$ & 0.5510 & 0.4171 & $\mathbf{0.6014}$ & $\mathbf{0.6769}$ & $\mathbf{0.3506}$ & $\mathbf{0.5538}$ \\
\bottomrule
\end{tabular}
}
\end{table}


Now, we focus on the MRR metric. As shown in Table \ref{tab:1-mrr}, the SelfAtt model performs the best in four languages, and NBoW has the highest MRR scores in two languages. NBoW performs slightly better than RNN even though they have very close MRR scores. CNN is the worst model when in most languages. This finding indicates that SelfAtt, NBoW, and RNN are more powerful as encoders than CNN.

Based on the comparisons of languages, the models perform better on Go, then Python, but have the worst performance on Ruby, then JavaScript. This finding is consistent with similar results from the literature \cite{Husain2019CodeSearchNetCE}.


\begin{table}[thb]
\centering
\caption{RQ1: NDCG Scores of Baseline Models}
~\\
\label{tab:1-ndcg}
\resizebox{\linewidth}{!}{
\begin{tabular}{lccccccc}
\toprule
Encoder & Go & Java & JavaScript & PHP & Python & Ruby & Avg. \\
\midrule
NBoW & $\mathbf{0.1165}$ & $\mathbf{0.1989}$ & $\mathbf{0.0653}$ & $\mathbf{0.1494}$ & $\mathbf{0.2994}$ & $\mathbf{0.1294}$ & $\mathbf{0.1598}$ \\
1D-CNN & 0.0139 & 0.1163 & 0.0098 & 0.1238 & 0.2044 & 0.0395 & 0.0846 \\
Bi-RNN & 0.0311 & 0.1220 & 0.0253 & 0.0976 & 0.1845 & 0.0552 & 0.0860 \\
SelfAtt & 0.0367 & 0.0951 & 0.0261 & 0.0785 & 0.1394 & 0.1152 & 0.0818 \\
\bottomrule
\end{tabular}
}
\end{table}

Let us now focus on the NDCG metric. As detailed in Table \ref{tab:1-ndcg}, the NBoW model is the most powerful and it outperforms other models by a considerable margin. This finding supports that NBoW is a better encoder than the others. Meanwhile, we find that all baselines perform the best on Python and perform the worst on JavaScript, Go, and Ruby. If we relate this finding to that of the MRR scores, we could see it is easy for models to have decent results on Python but much harder on JavaScript and Ruby. This phenomenon is explainable as follows: based on our manual observations on the corpus, the code snippets in Python generally have stronger readability, and on the contrary, the JavaScript and Ruby snippets tend to overuse meaningless one-character names and abstractive functional programming statements, reducing the ability of the neural network to grasp the meaning of variables.


According to all these results, it is clear that the NBoW model is the strongest baseline model for code search. The most obvious difference between NBoW and other baselines is that NBoW does not use the order information of the token sequences. Considering that, SelfAtt is the most competitive model which studies the context information.

\begin{tcolorbox}[fonttitle=\bfseries,title={Answer to Research Question 1}]
  \textbf{\rqaaa}
  \tcblower
  Overall, the NBoW model is the best concerning the learning and generalization ability. Among the models which consider some context information, the SelfAtt model performs the most competitively. According to the CodeSearchNet dataset, neural code search tends to be easier on Python but harder on JavaScript and Ruby.
\end{tcolorbox}

\subsection{Results of RQ2}

In this section, we measure the accuracy of unimodal learning based on the different tree representations introduced in \autoref{sec:approach}. We use the naming convention of Uni-Code and other short names explained in \autoref{subsec:mrq2}. 

To make the comparison between various representations more intuitive, we display the change ratio, that is, the relative increase or decrease degree of the MRR score relative to the Uni-code score. The change ratio is not a new metric.

\begin{table}[thb]
\centering
\caption{RQ2: MRR Scores of Unimodal Representations. Uni-LCRS can be considered the best representation.}
~\\
\label{tab:2-mrr}
\resizebox{\linewidth}{!}{
\begin{tabular}{lcccc}
\toprule
\multirow{2}{*}{Representation} &
\multicolumn{2}{c}{Python} &
\multicolumn{2}{c}{Ruby} \\
\cmidrule(lr){2-3}\cmidrule(lr){4-5}
 & Score & Change & Score & Change \\
\midrule
Uni-Code & 0.7533 & - & 0.3113 & - \\
\cmidrule{1-5}
Uni-RootPath & 0.8305 & 10.26\% & 0.3440 & 10.50\% \\
Uni-LeafPath & 0.6744 & -10.46\% & 0.2752 & -11.59\% \\
Uni-SBT & 0.8662 & 14.99\% & $\mathbf{0.3639}$ & $\mathbf{16.88\%}$ \\
Uni-LCRS & $\mathbf{0.8707}$ & $\mathbf{15.60\%}$ & 0.3423 & 9.96\% \\
\bottomrule
\end{tabular}
}
\end{table}


Table \ref{tab:2-mrr} gives the MRR scores and change ratios for the considered unimodal representations. The rows are for code representations and the columns are for languages. For example, the MRR score on Python for the Uni-RootPath representation is 0.8305 and its change ratio is 10.26\%.

As reflected in Table \ref{tab:2-mrr}, most tree-serialized representations are more effective than the token representation of Uni-Code. The best representation is Uni-SBT, for it has the best scores in both two languages.

The best representation on Python and Ruby is respectively Uni-LCRS and Uni-SBT, with an increase of 15.60\% and 16.88\% on MRR scores. Besides, Uni-LeafPath performs the worst. Even though not the best, Uni-RootPath still surpasses Uni-Code and brings stable improvements on both two languages. Uni-SBT has the best and most stable performance. In short, the traversal-based representations, like SBT and LCRS, perform better than the sampling-based representations, like RootPath and LeafPath. To our knowledge, this finding has never been reported in the literature.

\begin{table}[thb]
\centering
\caption{Coverage Ratios of Unimodal Representations}
~\\
\label{tab:2-ratio}
\resizebox{\linewidth}{!}{
\begin{tabular}{lcccc}
\toprule
\multirow{2}{*}{Representation} &
\multicolumn{2}{c}{Python} &
\multicolumn{2}{c}{Ruby} \\
\cmidrule(lr){2-3}\cmidrule(lr){4-5}
 & Link & Node & Link & Node \\
\midrule
Uni-Code & 0\% & 100\% & 0\% & 100\% \\
\cmidrule{1-5}
Uni-RootPath & 25.02\% & 67.22\% & 27.35\% & 67.59\% \\
Uni-LeafPath & 2.06\% & 11.48\% & 8.53\% & 30.51\% \\
Uni-SBT & 100\% & 100\% & 100\% & 100\% \\
Uni-LCRS & 10.91\% & 100\% & 9.15\% & 100\% \\
\bottomrule
\end{tabular}
}
\end{table}

The link coverage and node coverage, defined in \autoref{subsec:metric}, help explain why traversal-based representations are better. Table \ref{tab:2-ratio} shows the link coverage and node coverage, and we see that Uni-SBT and Uni-LCRS representations usually preserve more semantic and syntactic information than others. It also explains why tree-serialized representation brings improvements over Uni-Code (Uni-Code contains no syntactic information at all as the link coverage is zero).

Compared with Uni-Code, the Uni-RootPath representation contains more syntactic information even though less semantic information, because of its higher link coverage but lower node coverage. This explains why Uni-RootPath has higher MRR scores on Python and Ruby. However, even though Uni-LeafPath has more syntactic information as well, its link coverage is not so high. It explains the reason why Uni-LeafPath obtains lower MRR scores.

From the perspective of link coverage and node coverage, Uni-SBT is the most competitive representation because its value of 100\% is the highest. Another traversal-based representation, Uni-LCRS, obtains close but not so stable MRR scores because its link coverage is much smaller. However, if we compare with Uni-Code, then it is obvious that the syntactic information conveyed by Uni-LCRS is important. It proves that syntactic information is beneficial, meanwhile, enough syntactic information could improve the results stably.

Traversal-based representations contain more information, revealed by link coverage and node coverage, so they perform better than the sampling-based representations. For the path length specified in sampling and filtering operations, it is reasonable for Uni-RootPath to have such coverages. Besides, a leafpath is a subpath merged by a pair of rootpath, so a set of leafpath is more likely to have duplicated links and nodes. Thanks to the traversal strategy, Uni-SBT and Uni-LCRS have full node coverage. Unlike SBT, because of the transformation from binary tree to arbitrary tree, LCRS only reserves the links between the parent node and its most left child node.


\begin{tcolorbox}[fonttitle=\bfseries,title={Answer to Research Question 2}]
  \textbf{\rqbbb}
  \tcblower
  For unimodal learning, the tree-serialized representations are better than plain text representations. Overall, Uni-SBT is the best tree-serialized representation.
  Our results also shows that traversal-based representations tend to be better than sampling-based representations.
\end{tcolorbox}

\subsection{Results of RQ3}

In this section, we measure the accuracy of tree-serialized representations with the multimodal learning model. Recall the naming of Uni-Code introduced in \autoref{subsec:mrq2} and other namings introduced in \autoref{subsec:mrq3}, for the intuitive comparison with Uni-Code, we display change ratios as well.

\begin{table}[thb]
\centering
\caption{MRR Scores of Multimodal Representations}
~\\
\label{tab:3-mrr}
\resizebox{\linewidth}{!}{
\begin{tabular}{lcccc}
\toprule
\multirow{2}{*}{Representation} &
\multicolumn{2}{c}{Python} &
\multicolumn{2}{c}{Ruby} \\
\cmidrule(lr){2-3}\cmidrule(lr){4-5}
 & Score & Change & Score & Change \\
\midrule
Uni-Code & 0.7533 & - & 0.3113 & - \\
\cmidrule{1-5}
Multi-RootPath & 0.8410 & 11.65\% & $\mathbf{0.3602}$ & $\mathbf{15.70\%}$ \\
Multi-LeafPath & 0.7926 & 5.23\% & 0.3307 & 6.23\% \\
Multi-SBT & 0.8536 & 13.32\% & 0.3400 & 9.22\% \\
Multi-LCRS & $\mathbf{0.8563}$ & $\mathbf{13.68\%}$ & 0.3470 & 11.45\% \\
\bottomrule
\end{tabular}
}
\end{table}

Table \ref{tab:3-mrr} gives the MRR scores and change ratios for different multimodal representations. The rows are for representations and the columns are for languages. For example, the MRR score on Python for the Multi-RootPath representation is 0.8410 and its change ratio is 11.65\%.

As reflected in Table \ref{tab:3-mrr}, Uni-Code performs always worse than multimodal representations, on both Python and Ruby. It clearly shows the advantage of our multimodal learning way, therefore, the combinations of Uni-Code and tree-serialized representations always bring improvements over Uni-Code.

Compared with others, multi-LeafPath obtains the worst performance. On Python and Ruby, multi-LCRS and multi-RootPath separately perform the best. Besides, multi-SBT has close MRR scores with theirs.

By comparing multimodal representations with their corresponding unimodal representations, we find that sampling-based representations obtain higher MRR scores. The biggest increase is on LeafPath even though both Uni-LeafPath and Multi-LeafPath are not so good as others. The change ranges from Uni-LeafPath to Multi-LeafPath on Python and Ruby are 15.69\% and 17.82\%. However, the MRR scores of traversal-based representations become slightly worse.

\begin{table}[thb]
\centering
\caption{Coverage Ratios of Multimodal Representations}
~\\
\label{tab:3-ratio}
\resizebox{\linewidth}{!}{
\begin{tabular}{lcccc}
\toprule
\multirow{2}{*}{Representation} &
\multicolumn{2}{c}{Python} &
\multicolumn{2}{c}{Ruby} \\
\cmidrule(lr){2-3}\cmidrule(lr){4-5}
 & Link & Node & Link & Node \\
\midrule
Uni-Code & 0\% & 100\% & 0\% & 100\% \\
\cmidrule{1-5}
Multi-RootPath & 25.02\% & 100\% & 27.35\% & 100\% \\
Multi-LeafPath & 2.06\% & 100\% & 8.53\% & 100\% \\
Multi-SBT & 100\% & 100\% & 100\% & 100\% \\
Multi-LCRS & 10.91\% & 100\% & 9.15\% & 100\% \\
\bottomrule
\end{tabular}
}
\end{table}

Because of the combinations between the tree-serialized representations and the token representation, as shown in Table \ref{tab:3-ratio}, all multimodal representations have full node coverages. Therefore, only the link coverage, namely the completeness of syntactic information, introduces the difference in the MRR scores. It is reasonable for Multi-LeafPtah to perform badly because its link coverage is always the lowest, and that also explains why the MRR scores increase dramatically from Uni-LeafPath to Multi-LeafPath.

Even though Multi-SBT has the highest link coverage, its scores are not always as good as Multi-RootPath or Multi-LCRS. It indicates that the combination of Uni-SBT and Uni-Code does not necessarily help reveal more information. The cause is that Uni-SBT representation has as complete semantic information as Uni-Code, and meanwhile more syntactic information. Therefore, the combination with the Uni-Code representation is not capable to complement more information, but instead, adds some noise to the syntactic information. Similarly, it to some extent explains why Uni-SBT is better than Multi-SBT. Actually, the same observation is observed in Multi-LCRS as well, on Python but not on Ruby. That it is not obvious in Multi-LCRS is because the link coverage of Uni-LCRS is a bit low, so it could be affected by the random syntactic information of Uni-Code.



\begin{tcolorbox}[fonttitle=\bfseries,title={Answer to Research Question 3}]
  \textbf{\rqccc}
  \tcblower
  The multimodal representations surpass the unimodal ones for code search. Considering conciseness and effectiveness, RootPath and SBT are the recommended sampling-based and traversal-based representations respectively.
  Overall, a multimodal representation can be considered as the new state-of-the-art for neural code search.
\end{tcolorbox}

\subsection{Threats to Validity}
\label{sec:discussion}

The multimodal representation approach and uni-model representation approach only experiment on two languages. According to our results, the accuracy scores reveal slight differences in different languages, but we are not sure whether it is caused by the data amounts or the data characteristics. It should make our conclusions more reliable if there are adequate experiments on other languages.

The query data are not necessarily of high quality or most semantically similar to respective code snippets. The whole corpus is using documentation texts in code snippets as the query data. However, documentation texts are likely to be outdated or off-topic, sometimes these texts are even automatically generated. Besides, the code-query pairs from the corpus rely on the hypothesis that code and query data from the same pair is most semantically similar. However, this hypothesis could be affected by the quality of the corpus.

\section{Related Work}
\label{sec:related}

There have been lots of work studying code search \cite{Sadowski2015HowDS,Liu2020OpportunitiesAC}. Limited to source code search, the typical application scenarios and progress of code search are as follows.

\subsection{API-based Code Search}

API tokens and call sequences usually contain valuable semantic information. Some research is centered on the cluster and search problem of similar code based on the API data.

MAPO \cite{Zhong2009MAPOMA} extracts API call sequences from code snippets and then groups them into clusters. The most frequent call sequences are recognized as usage patterns for further recommendation. Instead of the invocation sequences of API functions, eXoaDocs \cite{Kim2010TowardsAI} approximates the semantic feature vectors for clustering. GrouMiner \cite{Nguyen2009GraphbasedMO} mines API usage patterns by representing code as graphs to utilize the structural information. UPMiner \cite{Wang2013MiningSA} introduces two quality metrics and designs a two-step clustering strategy to mine succinct and high-coverage API usage patterns. MUSE \cite{Moreno2015HowCI} combines program slicing and text-based clone detection to rank the groups of similar examples, considering their popularity. CodeKernel \cite{Gu2019CodeKernelAG} represents code samples as object usage graphs, and then uses a graph kernel to embed them into a continuous space for clustering, and eventually rank code samples considering both centrality and specificity.

\subsection{Constraint-based Code Search}

The input-output pairs and control flow graph of programs are capable to generate program constraints \cite{Stolee2014SolvingTS}. They specify what behaviors the target code snippets are expected to have.

Satsy \cite{Stolee2016CodeSW} utilizes symbolic execution engines to support the encoding work on multi-path programs, and then ranks code candidates based on the matching degree between the query and encoded program paths. Quebio \cite{Chen2020EnhancingEC} encodes code snippets into path constraints via symbolic analysis and returns the fulfilled programs. Compared with Satsy, Quebio supports more data types and operations, and the invocation to library APIs. When structural or semantic properties of code snippets are expressed as logic facts, ALICE \cite{Sivaraman2019ActiveIL} extracts a logic query from the given code example and infers new logic queries based on feedback. Yogo \cite{Premtoon2020SemanticCS} studies the dataflow equivalences with rewrite rules, to recognize equivalent code fragments as long as they are the same higher-level concept.

\subsection{Example-based Code Search}

Given a code example, similar code snippets are retrieved at varying semantic and syntactic levels. It is common to find code snippets that are similar but subtly different as results.

SourcererCC \cite{Sajnani2016SourcererCCSC} computes the similarity of code snippets based on the overlap degree in token level. FaCoY \cite{Kim2018FaCoYA} analyzes the functionality of the code example and utilizes representative tokens from available code snippets which own that functionality, to find the most similar candidates. CodeNuance analyzes code commonalities and differences to support exploratory code search \cite{Liu2018SupportingEC}. Aroma \cite{Luan2018AromaCR} first assembles code snippets that match the query, then computes the similarity based on their structural features, and eventually clusters and intersects relevant code candidates. Siamese \cite{Ragkhitwetsagul2019SiameseSA} improves code search by transforming code into a multi-representation and reducing the long query to the query composed of only rare tokens. It computes a customized score for each code snippet to support incremental updates to the source code bases.

\subsection{Text-based Code Search}

In text-based code search, the query intentions are expressed in natural language. It is feasible to see code data as text, so information retrieval techniques apply. Besides, deep learning methods show superiority in capturing data similarity.

Sourcerer \cite{Linstead2008SourcererMA} simply regards the code samples as text and then computes the TF-IDF \cite{Haiduc2013AutomaticQR} score to measure the relevance level with the given query. RACS \cite{Li2016RelationshipawareCS} extracts action relationship graph from the query data and represents API usage patterns from the collected code snippets as the method dependency graphs. In this way, the code search problem is reduced to the problem of finding similar method dependency graphs for a given action relationship graph. CodeExchange and CodeLikeThis are capable to leverage the results in query formulation. The former lists characteristics to search similar results for new queries and the latter directly searches results analogous to current results \cite{Martie2017UnderstandingTI}. By utilizing the API documentation, query words could be expanded \cite{Lv2015CodeHowEC,Zhang2018ExpandingQF} with the potential APIs referred to. To improve the matching degree between the query and code data, queries are studied to be well reformulated by considering the crowd knowledge and other techniques such as context awareness \cite{Sirres2017AugmentingAS,Rahman2018EffectiveRO,Rahman2019AutomaticQR,Huang2019EnhanceCS}.

Since the code search task is being explained as a probabilistic model \cite{Allamanis2015BimodalMO}, various deep learning solutions are proposed. CODE-NN \cite{Iyer2016SummarizingSC} leverages LSTM to build the translation model between code snippets and natural language texts. NCS \cite{Sachdev2018RetrievalOS} combines word embedding, TF–IDF weighting, and a supervision layer for code search. UNIF \cite{Cambronero2019WhenDL} extends NCS by using the bag-of-words model and turn to use the attention mechanism for embedding weights. SCS \cite{Husain_2018,GitHub_Blog_2018} uses sequence-to-sequence networks to map code and query data into the shared vector space, to make semantically similar code-query pairs close to each other. Instead of merely regarding both code and query as text sequence, DeepCS \cite{Gu2018DeepCS} extracts function names, invocation sequences, and token set from the code data, and compute the cosine similarity of code and query vectors to support the search process. MMAN \cite{Wan2019MultimodalAN} utilizes the data of multiple modalities for better code representation. It represents tokens, abstract syntax tree, and control flow graph by using LSTM, Tree-LSTM \cite{Tai2015ImprovedSR}, and GGNN \cite{Li2016GatedGS} respectively. Aimed to tackle certain issues in code search, CQIL \cite{Li2020LearningCI} models the semantic correlations between code and query with hybrid representations. Similarly, NJACS \cite{Hu2020NeuralJA}, CARLCS \cite{Shuai2020ImprovingCS}, TabCS \cite{Xu2021TwoStageAM} and SANCS \cite{Fang2021SelfAttentionNF} learn attention-based representations of code and query with the co-attention mechanism.

\section{Conclusion}
\label{sec:conclusion}

Our work is about applying tree-serialized representations by the multimodal learning way for code search. The core idea of our multimodal representation model is to utilize both semantic and syntactic information of code snippets.

We designed a novel tree structure named Simplified Semantic Tree (SST). SST is more semantically informative than AST, then we introduced several tree-serialization methods on SST to build tree-serialized representations to complement the token representation. Moreover, we combined tree-serialized representations with the token representation as multimodal representations. Our multimodal learning model follows the pseudo-siamese network architecture and adopts the SelfAtt model as its encoders. Last, we defined two intuitive quantification metrics to quantify the completeness of the semantic information and syntactic information conveyed by token representation and tree-serialized presentations.

Based on our experiments on the large-scale multi-language corpus, the SelfAtt model is most satisfying considering the context information. Among all the introduced tree-serialized representations, traversal-based representations usually perform better than the sampling-based ones. More results suggest that tree-traversal representations bring improvements of at most 16.88\% and the multimodal model raises the MRR scores by at most 17.82\%. We recommend adopting tree-serialized representations as well as the multimodal learning model for code search, whenever possible.

In the future, we will explore other software engineering problems where our work might apply. Besides, we are to investigate more aspects of the side of code data processing, for example, study the design details of existing novel designs similar to SST and seek potential theoretical support.



\section*{Acknowledgment}

This work was partially supported by the Wallenberg AI, Autonomous Systems and Software Program (WASP) funded by Knut and Alice Wallenberg Foundation, and by the Swedish Foundation for Strategic Research (SSF), with computational resources provided by the Swedish National Infrastructure for Computing.

\newpage

\printbibliography[heading=bibintoc]

\end{document}